\documentclass[11pt, a4paper]{article}

\usepackage[english]{babel}
\usepackage[utf8]{inputenc}
\usepackage[T1]{fontenc}

\usepackage{tikz, pgfplots}
\usetikzlibrary{shapes, arrows, positioning}

\usepackage{amsmath, amssymb, amsfonts, amsthm}

\usepackage{xcolor, graphicx}

\usepackage{caption, subcaption}

\usepackage{booktabs}

\usepackage{float}
\usepackage{geometry}
\usepackage{multirow}
\usepackage{siunitx}
\usepackage{enumitem}
\usepackage{gensymb} 
\usepackage{siunitx} 
\usepackage{authblk}

\usepackage[numbers,super]{natbib}

\definecolor{pastelblue}{rgb}{0.0, 0.5, 1.0} 
\usepackage[colorlinks=true, citecolor=pastelblue, linkcolor=blue, urlcolor=blue]{hyperref}

\makeatletter
\renewcommand{\@cite}[2]{%
  \textsuperscript{\Large\textbf{\textcolor{pastelblue}{#1}}}%
}
\makeatother

\usepackage{titlesec}

\titlespacing*{\section}{0pt}{1ex plus 0.5ex minus .2ex}{0.5ex plus .1ex}

\titlespacing*{\subsection}{0pt}{3ex plus 0.3ex minus .1ex}{0.5ex plus .1ex}
\DeclareCaptionLabelFormat{boldpipe}{\textbf{#1~#2|}\ }

\title{Dielectric Nanotomography Based on Electrostatic Force Microscopy: a Numerical Analysis.}

\author[1]{Rene Fabregas}
\author[1,2]{Gabriel Gomila}
\affil[1]{Institut de Bioenginyeria de Catalunya (IBEC), The Barcelona Institute of Science and Technology (BIST), c/ Baldiri i Reixac 11-15, 08028, Barcelona, Spain}
\affil[2]{Departament d'Enginyeria Electrònica i Biomèdica, Universitat de Barcelona, c/ Martí i Franqués 1, 08028, Barcelona, Spain}
\date{}

\begin{document}

\maketitle

\begin{abstract}
Electrostatic Force Microscopy (EFM) has demonstrated the capability to image nanoscale objects buried below the surface. Here, we show theoretically that this capability can be used to obtain nanotomographic information, i.e. physical dimensions and dielectric properties, of buried nano-objects. These results constitute a first step towards implementing a non-destructive dielectric nanotomagraphy technique based on EFM with applications in Materials and Life Sciences.
\end{abstract}


\section{Introduction.}
3D nanoscale tomographic imaging techniques have experienced an enormous demand in recent years due to the increasing needs of Nanotechnology and their applications in Electronics and Material Science, as well as, by the requirement of a higher spatial resolution imaging in Molecular and Cell Biology on intact samples. Examples of application of 3D nanoscale tomography include modern 3D stacked microelectronic devices, nanocomposite materials, and nanotoxicity and drug delivery studies\citep{soliman2017}.

Currently, the gold standards for 3D nanoscale tomographic imaging are Electron Microscopy and X-Ray Microscopy, with specific adaptations for Materials\citep{mobus2007,withers2007,midgley2009} and Life Sciences\citep{le2005,larabell2010,milne2009,subramaniam2007,lucic2005}. However, these nanoscale tomographic techniques still face some limitations when applied to some samples of relevant applied interest, specially in the Life Science realm. Examples include the tomographic imaging of nanoengineered materials within soft matrices and living cells relevant for Health, Food and Biomedical studies (including drug delivery, nanotoxicity and infectious processes). In addition, existing 3D nanoscale tomographic techniques also show limitations to map the physical properties of buried nanomaterials (e.g. mechanical, electrical or magnetic), which are relevant in many areas of Science and Technology (a relevant exception to that is Electron Holography\citep{midgley2009}). Finally, some of the existing nanotomographic techniques require aggressive sample preparation methods and imaging conditions, which include flash-frozen and imaging at cryogenic temperatures for biological samples\citep{lucic2005}, or sample sectioning and milling for some solid-state materials applications\cite{mobus2007}.

To cope with some of these limitations alternative nanoscale tomographic techniques based on Scanning Probe Microscopy (SPM) are under investigation. The first approaches were based on the use of correlative microscopy methods (correlative AFM nanotomography)\citep{magerle2000,sperschneider2010}. However, these methods are still destructive and limited by the sectioning or milling process of the sample. Later, the focus was directed towards non-destructive nanotomographic SPM techniques, by exploiting the capability of some SPM techniques to access the sub-surface properties of intact samples. Examples include Scanning Near Field Ultrasound Holography\citep{shekhawat2005,tetard2008}, Mode Synthesizing Atomic Force Microscopy\citep{tetard2010}, Electrostatic Force Microscopy\citep{jespersen2007}, Scanning Microwave Impedance Microscopy\citep{yang2012} and Scanning Near Field Optical Microscopy\citep{govyadinov2014}. These SPM techniques share in common the capability to probe relatively large sub-surface volumes of the sample, where they can detect variations in the sub-surface physical properties (e.g. mechanical, electrical, electromagnetic or optical properties) with nanoscale spatial resolution\citep{bonnell2012}. These SPM techniques, besides being non-destructive, offer the advantage of being applicable virtually to any sample (soft and stiff, conductor and dielectric) and to any environmental condition (including ambient and liquid environments). However, until now, the nanotomographic capabilities of these SPM sub-surface techniques, i.e. the possibility to reconstruct nanoscale 3D tomographic images from the subsurface 2D SPM images has not been demonstrated in most cases\citep{govyadinov2014}.  

In the present work, we precisely address this issue, and analyze the nanotomographic capabilities of Electrostatic Force Microscopy (EFM). EFM is of one of the sub-surface SPM techniques that has received more attention\citep{jespersen2007,takano2001,zhao2010,riedel2011,arinero2012,alekseev2012,cadena2013,thompson2013,castaneda2015,patel2016,takano2000a} but still its tomographic reconstruction capabilities remain unclear.  We show by means of numerical calculations that EFM can provide nanotomographic information of nano-objects buried below the sample surface under specific conditions and discuss its practical implementation. Present results show that EFM is a good candidate to become a wide use non-destructive SPM nanotomographic technique for applications in Materials and Life Sciences.

\section{Results.}

\subsection{System under study.}

\begin{figure}[H]
    \centering
    \includegraphics[width=0.45\textwidth]{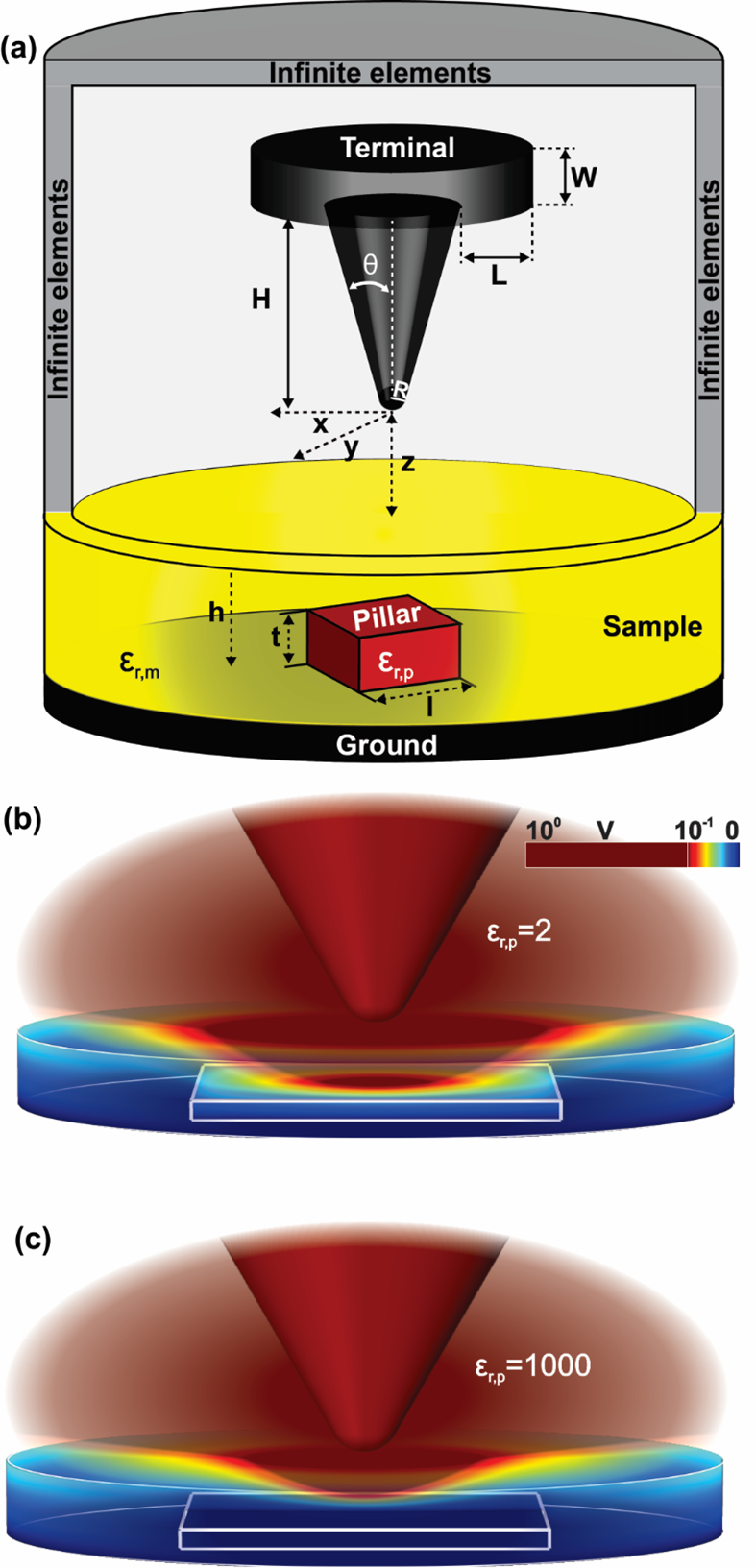}
    \caption{(a) Schematic representation of the EFM buried nanopillar system under study. (b) and (c) Electric potential distributions corresponding to a buried pillar with $\varepsilon_{r,p} =$ 2 and 1000, respectively, and with thickness $t = \SI{50}{nm}$ and side length $l = \SI{1000}{nm}$. The remaining parameters are: $R = \SI{100}{nm}$, $\theta = 25^{\circ}$, $H = \SI{12.5}{\micro m}$, $L = \SI{0}{nm}$, $W = \SI{3}{\micro m}$, $h = \SI{200}{nm}$, $\varepsilon_{r,m} = 4$, and tip-sample distance $z = \SI{20}{nm}$.
}
    \label{fig:fig1a}
\end{figure}

We investigate theoretically the nanotomographic capabilities of EFM for the system schematically shown in \textbf{Figure~\ref{fig:fig1a}}. The system consists of a thin dielectric film matrix of thickness $h$, and dielectric constant $\varepsilon_{r,m}$, containing a buried square nano-parallelepiped of thickness $t$, and side length $l$, and relative dielectric constant $\varepsilon_{r,p}$. The nanopillar is assumed to be supported on a metallic substrate. The sample is assumed to be imaged by an EFM metallic tip consisting of a cone ended with a tangent sphere, as in previous works\citep{gomila2014} (see Materials and Methods). \textbf{Figures 1b} and \textbf{1c} show examples of numerically calculated electric potential distributions for the case of a buried pillar of thickness $t=50$ nm and side length $l=1000$ nm, and dielectric constants $\varepsilon_{r,p}=2$ and 1000, respectively, buried in a dielectric matrix 200 nm thick and with dielectric constant $\varepsilon_{r,m}=4$ (see Supplementary Information (SI) for some additional images and animations).

\subsection{Forward problem.}
EFM images can be numerically calculated when all parameters are known by computing the electric force acting on the tip in response to a voltage bias $V$ applied between the tip and the metallic substrate while scanning the tip laterally in the $X$ and $Y$ directions at a fixed distance $z$ from the sample surface (forward problem). The electric force acting on the tip is calculated by integrating the Maxwell stress tensor on the tip surface.

\textbf{Figures 2a-2c} (res. \textbf{Figs. 2d-2f}) show calculated EFM images for the system in \textbf{Fig. 1a} for buried structures with side lengths $l = \SI{1}{mm}$, \SI{500}{nm} and \SI{100}{nm}, respectively, and relative dielectric constant $\varepsilon_{r,p} = 2$ (res. $\varepsilon_{r,p} = 10$). In all cases, the thickness of the buried structure is $t = \SI{50}{nm}$, while the thin film matrix thickness is $h = \SI{100}{nm}$ and its relative dielectric constant, $\varepsilon_{r,m} = 4$. The images have been calculated for a tip with radius $R = \SI{100}{nm}$, cone half angle $\theta = 25^{\circ}$, cone height $H = \SI{12.5}{mm}$, cantilever thickness $W = \SI{3}{mm}$ and cantilever length $L = \SI{0}{nm}$ (for thin films the cantilever effects are negligible\citep{gramse2012}). The tip-sample distance is $z = \SI{20}{nm}$ in all calculated images. The EFM images are represented in terms of the capacitance gradient 
$\frac{\partial^2 C}{\partial z^2} = -\frac{1}{2} \frac{\partial F_{el}}{\partial V^2}$, 
where $F_{el}$ is the calculated electric force acting on the tip and $V$ is the applied voltage (see Materials and Methods). The calculated EFM images clearly display the presence of the buried pillar and confirm the subsurface imaging capabilities of EFM. The contrast in the images is negative (res. positive) when the dielectric constant of the buried structure is lower (res. larger) than that of the surrounding thin film matrix. We note the presence of tip convolution effects in the calculated EFM images, which can be directly appreciated by comparing the distribution of the contrast in the EFM images with the physical dimensions of the buried structure shown by the white dashed lines in \textbf{Figs. 2a-2f}. Tip broadening effects in EFM images follow different rules than the tip broadening effects in topographic imaging, due to the long-range nature of the electrostatic interaction\citep{gomez2001}.

The main consequences of EFM tip broadening for nanotomographic applications are that the lateral physical dimensions of the buried object can only be determined approximately from the EFM images, and that the actual shape of the buried object can be difficult to identify, especially when the side lengths are smaller than the tip radius (see \textbf{Figs. 2c} and \textbf{2f}).

\textbf{Figures 2g-2i} show cross-section (absolute) capacitance gradient profiles obtained from the calculated EFM images at different tip sample distances. In these figures, and in the following ones, the dashed lines correspond to $\varepsilon_{r,p} = 2$ while solid lines to $\varepsilon_{r,p} = 10$. The capacitance gradient values, and the image contrast, decrease quickly as the tip sample distance increases, as expected. This fact is further illustrated in \textbf{Fig. 2j} where we plot the maximum EFM capacitance gradient contrast as a function of the tip sample distance, for pillars with different side lengths and different dielectric constants. In addition to the tip sample distance, the capacitance gradient contrast in a sub-surface EFM image depends, on the dimensions and dielectric properties of the matrix and buried object. In particular, it depends on: (i) the lateral size of the buried pillar: the contrast increases with the later size until it saturates to a constant value independent from the side length, for side lengths larger than a few times the tip radius (see \textbf{Fig. 2k}); (ii) the thickness of the buried pillar: for a given matrix thickness the contrast increases until the thickness of the buried structure approaches the matrix thickness (\textbf{Fig. 2l}); (iii) the dielectric constant of the buried pillar: the contrast increases roughly logarithmically with $\varepsilon_{r,p}$ up to $\varepsilon_{r,p} \sim 80-100$ where it tends to saturate (\textbf{Fig. 2m}); (iv) the thickness of the matrix: the contrast decreases by increasing the thickness of the matrix until it becomes undetectable for thicknesses larger than a few times the tip radius (\textbf{Fig. 2n}); and (v) the dielectric constant of the matrix: the contrast shows a maximum (res. minimum) for $2 < \varepsilon_{r,m} < 10$ (res. $10 < \varepsilon_{r,m} < 80$) and $\varepsilon_{r,p} > 10$ (res. $\varepsilon_{r,p} < 10$), and then tends to zero when the dielectric constant of the matrix is large $\varepsilon_{r,p} > 100$ (\textbf{Fig. 2o}). Finally, the capacitance gradient contrast also depends strongly on the tip geometry and dimensions, especially on the tip radius: it increase non-linearly as the tip radius increases, as shown in \textbf{Figs. 2p-2r} where the capacitance gradient contrast as a function of tip radius is plotted for different tip sample distances, and for three side-lengths of the buried pillar, $l = \SI{100}{nm}$, $\SI{500}{nm}$ and $\SI{1}{\mu m}$, respectively.
\begin{figure}[H]
    \centering
    \includegraphics[width=\textwidth]{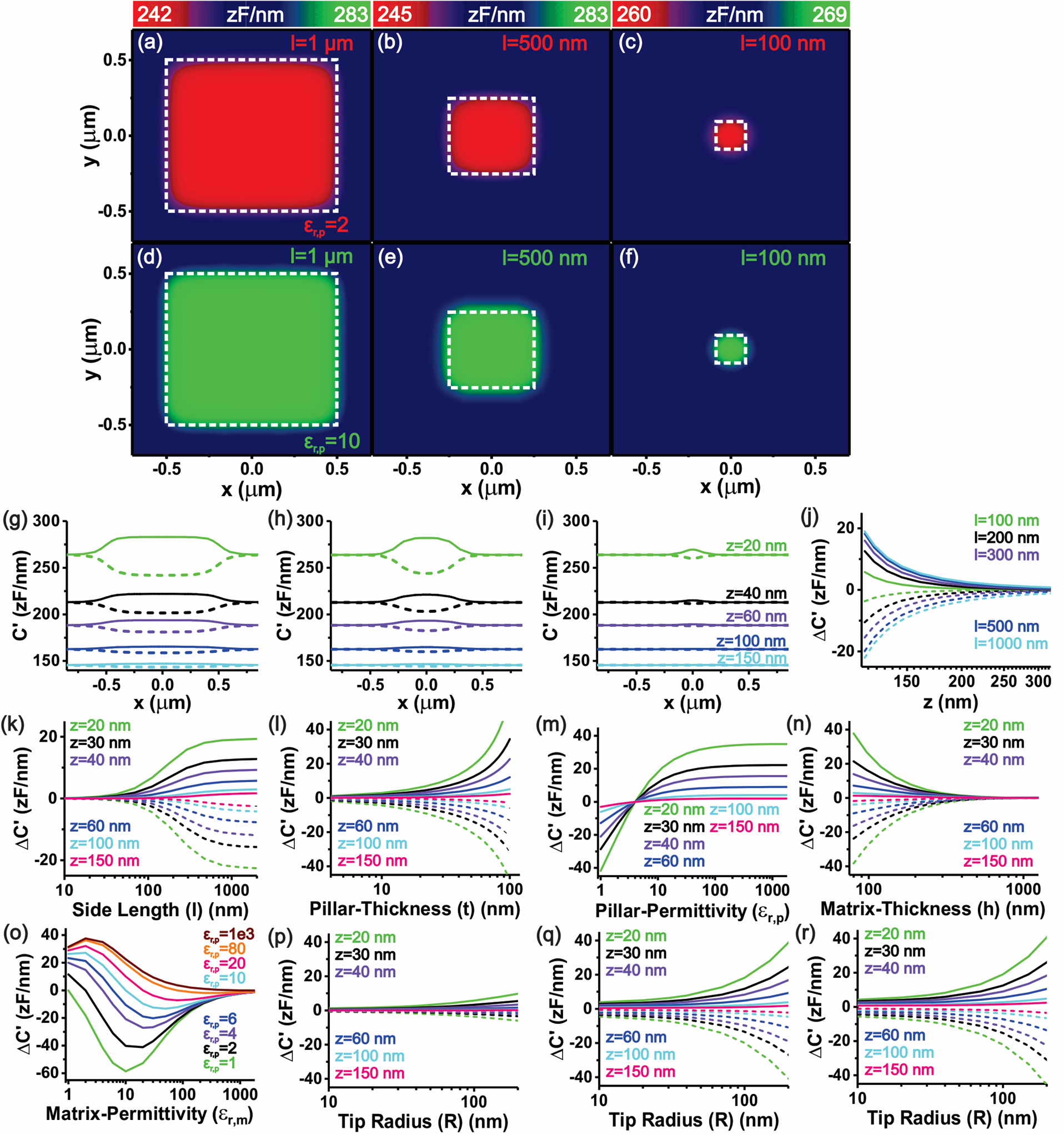}
    \caption{\textbf{(a)-(c)} Numerically calculated EFM images at tip-sample distance $z = \SI{20}{nm}$ of three different buried structures of thickness $t = \SI{50}{nm}$, dielectric constant $\varepsilon_{r,p} = 2$, and side lengths $l = \SI{1}{\mu m}$, \SI{500}{nm} and \SI{100}{nm}, respectively. Parameters of the tip and thin dielectric film matrix if not otherwise state: $R = \SI{100}{nm}$, $H = \SI{12.5}{\mu m}$, $L = \SI{0}{nm}$, $W = \SI{3}{mm}$; $h = \SI{100}{nm}$, $\varepsilon_{r,m} = 4$. \textbf{(d)-(f)} Idem but for $\varepsilon_{r,p} = 10$. \textbf{(g)-(i)} Capacitance gradient cross-section profiles along the center of the calculated EFM images shown in \textbf{(a)-(f)} for five tip-sample distances: $z = \SI{40}{nm}$ (black lines), \SI{60}{nm} (violet lines), \SI{100}{nm} (blue lines) and \SI{150}{nm} (cyan lines). Solid lines (res. dashed lines) correspond to the buried structures with $\varepsilon_{r,p} = 10$ (res. $\varepsilon_{r,p} = 2$). \textbf{(j)} Maximum capacitance gradient contrast (at the center of the buried structure) as a function of tip-surface distance for buried structures with side lengths $l = \SI{100}{nm}$, \SI{200}{nm}, \SI{300}{nm}, \SI{500}{nm} and \SI{1}{\mu m} with dielectric constants $\varepsilon_{r,p} = 2$ (dashed lines) and 10 (solid lines). \textbf{(k)} Contrast at the center of the buried structure as a function of its side length, $l$, for $\varepsilon_{r,p} = 2$ (dashed lines) and 10 (solid lines) and for six tip sample distances $z$. \textbf{(l)} Contrast at the center of the buried structure with side length $l = \SI{500}{nm}$ as a function of the thickness $t$ of the buried structure for six tip sample distances $z$, for $\varepsilon_{r,p} = 2$ (dashed lines) and 10 (solid lines). \textbf{(m)} Idem but as a function of the dielectric constant of the buried structure $\varepsilon_{r,p}$. \textbf{(n)} Idem but as a function of the thickness of the matrix $h$. \textbf{(o)} Contrast at the center of the buried structure with side length $l = \SI{500}{nm}$ as a function of the dielectric constant of the matrix $\varepsilon_{r,m}$ for tip sample distance $z = \SI{20}{nm}$. \textbf{(p)-(r)} Contrast at the center of the buried structure as function of tip radius $R$ for six different tip sample distances, $z$, and for buried structures with side length $l =\SI{100}{nm}$, \SI{500}{nm} and \SI{1}{\mu m}, respectively.}
    \label{fig:fig2a}
\end{figure}

\subsection{EFM tomographic reconstruction (inverse problem).}
Now we address the question of whether tomographic information on the buried pillar can be obtained from EFM images displaying subsurface information (inverse problem). As in most inverse problems, the main issue relates to whether a unique solution for the parameters can be found or not. Before answering this question, we first identify the unknowns to be determined. For the present EFM problem we can assume that the tip geometry, the thickness of the matrix $h$, and its dielectric constant $\varepsilon_{r,m}$, are known, since they can be determined from independent measurements\citep{gomila2014,fumagalli2014}. Therefore, we are left, in general, with the parameters defining the buried structure (shape, physical dimensions, dielectric constant) and its position within the thin film. To simplify further the problem, we assume that the shape and position of the object is known (a square parallelepiped with the base located on the metallic substrate). Due to the tip convolution effects, the shape of the buried object cannot always be inferred from the sub-surface EFM images, as we mentioned before. Moreover, we assume that the dielectric constant of the buried pillar is uniform, which is a reasonable assumption in most applications. According to these considerations, we are left with three unknowns, namely, the side length $l$, the thickness $t$, and the dielectric constant $\varepsilon_{r,p}$, of the buried pillar. Even with these simplifications and the reduction in the number of unknowns, their determination still constitutes a formidable inverse problem and illustrates faithfully the complexity of nanotomographic EFM reconstruction.

To determine the three unknowns, $\left(\varepsilon_{r,p},t,l\right)$, we have proposed the resolution of the optimization problem described by Eqs. (1)-(6) in the Materials and Methods section. The problem seeks to minimize the cumulative standard deviation $S$, between calculated and input EFM images (profiles) obtained at different tip-sample distances. Here, as input data we consider numerically calculated EFM images for given  $\left(\varepsilon^*_{r,p},t^*,l^*\right)$ (in practical applications, input data would be EFM images recorded at different tip--sample distances). Two examples of input EFM images at distances $z=20\,$nm and $z=60\,$nm are shown in \textbf{Figs. 3a} and \textbf{3b}, for the case of a buried pillar defined by  $\left(\varepsilon^*_{r,p},t^*,l^*\right)$=(10 nm, 50 nm, 1 $\mu$ m) within a dielectric matrix with $h=100\,$nm and $\varepsilon_{r,m}=4$, and imaged with a tip with $R=100\,$nm, $\theta=25^\circ$, $H=12.5\,\mu$m, $L=0\,$nm and $W=3\,\mu$m. \textbf{Figure 3c} shows capacitance gradient contrast cross-section profiles obtained from the input calculated EFM images at different tip--sample distances, $z= 20\,$nm, $30\,$nm, $40\,$nm, $60\,$nm and $100\,$nm.

To solve the optimization posed problem we have followed an algorithm that consists of different steps, which are described in \textbf{Figs. 3d-3l}. We first assume a given value for the side length $l$, and the tip-sample distance, $z$, and compute all possible capacitance gradient contrast values that can be obtained at the center of the buried structure and at the 1/3 of the full width (FW1/3M) (we chose the FW1/3M rather than the FWHM since it is more sensitive to the width of the pillar). To this end, we solve the forward problem by varying the values of the two unknown parameters, $\varepsilon_{r,p}$ and $t$. \textbf{Figures 3d} and \textbf{3e} show examples of the graphical representation of the calculated contrast capacitance gradient values as a function of $\varepsilon_{r,p}$ and $t$ for the sample described above for $l=991\,$nm and $z=20\,$nm. We then compare the possible capacitance gradient values with the corresponding input capacitance gradient values at the set distance, which in the example considered in \textbf{Fig. 3} (see \textbf{Fig. 3c}) are $\Delta C'_{in}(0) = \SI{19}{zF/nm}$ and $\Delta C'_{in}\left(x_{FW1/3M}\right) = \SI{6.4}{zF/nm}$ (these values are represented by surface planes in \textbf{Figs. 3d} and \textbf{3e}). The intersections of the two surfaces in \textbf{Figs. 3d} and \textbf{3e} give two sets of values for the couples $(\varepsilon_{r,p}, t)$ (black and orange curves in \textbf{Fig. 3f}). The intersection of these two sets of data gives the value of a possible solution to the problem for the given side length, $l$, and tip--sample distance, $z$, $\left( \varepsilon_{r,p}(l,z), t(l,z), l \right)$. For the example considered in \textbf{Fig. 3f}, corresponding to $l=991\,$nm and $z=20\,$nm, the intersection takes place at $\varepsilon_{r,p}=16.3$ and $t=40.6\,$nm (see zoom in \textbf{Fig. 3f}). This procedure is then repeated for all possible values of the side length, $l$, thus providing a family of possible solutions, for a given tip--sample distance, $z$, as shown in \textbf{Fig. 3g}.

\begin{figure}[htbp]
    \centering
    \includegraphics[width=\textwidth]{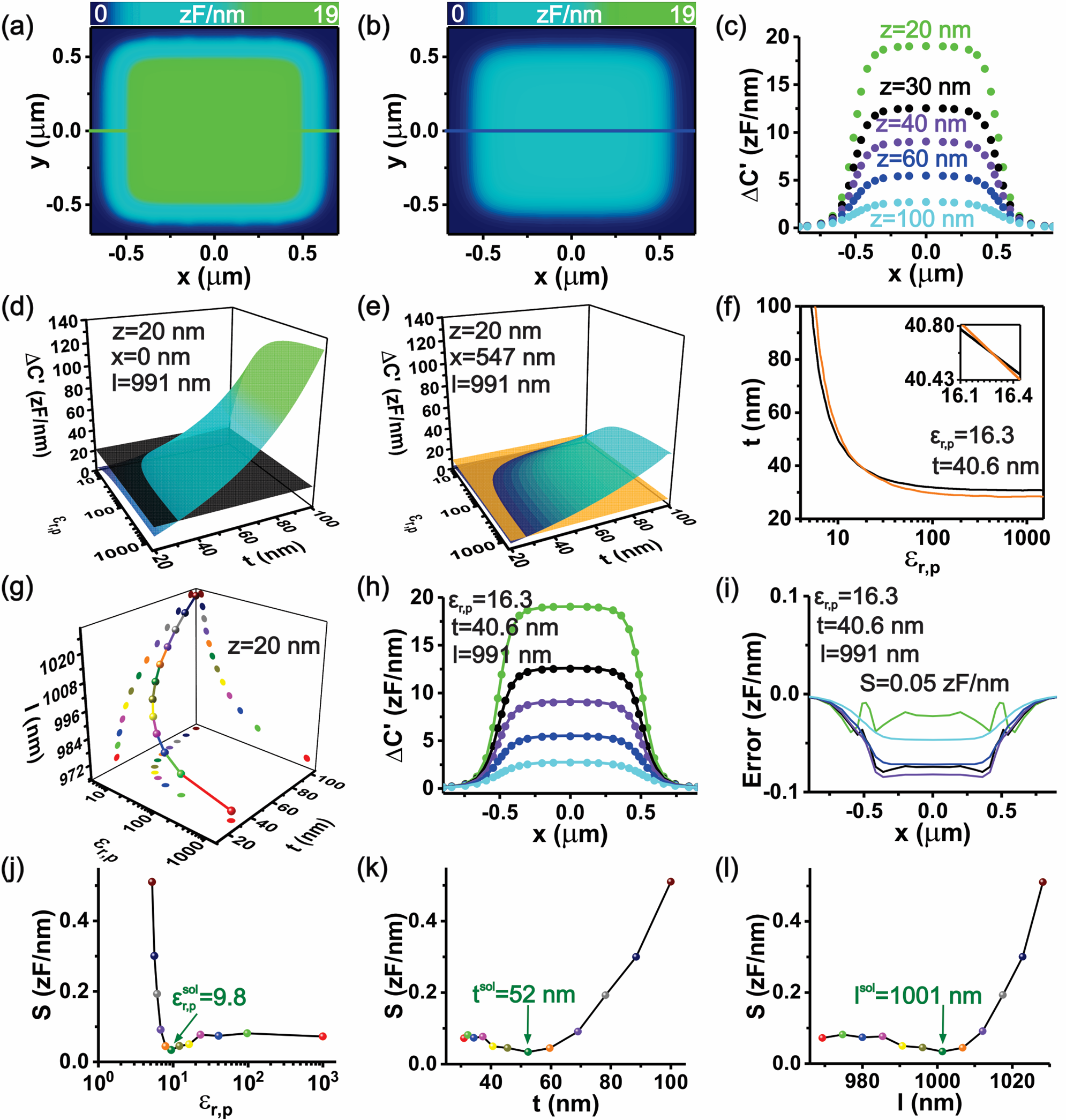}
    \caption{\textbf{(a)} and \textbf{(b)} Examples of input EFM images calculated at tip–sample distances of $z=20$~nm and $z=60$~nm, respectively. Parameters of the buried pillar: thickness $t^*=50$~nm, relative dielectric constant $\varepsilon_{r,p}^*=10$, and side length $l^*=1$~$\mu$m. \textbf{(c)} Capacitance gradient contrast profiles obtained from the input EFM images along the center for $z=20$~nm, $30$~nm, $40$~nm, $60$~nm, and $100$~nm. \textbf{(d)} Calculated possible values of capacitance gradient contrast at the center of the buried structure (green surface) as a function of $\varepsilon_{r,p}$ and $t$ for a tip–sample distance of $z=20$~nm and a buried pillar side length of $l=991$~nm. The gray plane corresponds to the contrast in the input capacitance gradient profile at the center of the buried structure at this distance. \textbf{(e)} Same as \textbf{(d)} but at the FW1/3M position. \textbf{(f)} Solution curves resulting from the intersection of the surfaces in \textbf{(d)} and \textbf{(e)}, respectively. \textit{Inset}: a zoom around the position where the curves intersect, yielding the possible solution values for $l=991$~nm and $z=20$~nm, $\varepsilon_{r,p}=16.3$, and $t=40.6$~nm. \textbf{(g)} Graph of the possible solutions for the set of parameters that accurately match the contrast at the center and at FW1/3M of the input profile for $z=20$~nm. \textbf{(h)} Comparison of the profiles calculated for one of the possible solutions in (f) (lines) with the input capacitance gradient profiles. (i) Capacitance gradient profile errors resulting from the comparison in \textbf{(h)}. The total error $S$ for this distance and possible solution is $S=0.05$~zF/nm. \textbf{(j)}, \textbf{(k)}, and \textbf{(l)} Representation of the total error $S$ as a function of the dielectric constant, thickness, and side length of the buried structure, respectively. The minima of these plots correspond to the solution of the problem, which in this case is $\left( \varepsilon^{sol}_{r,p}, t^{sol}, l^{sol} \right)=(9.8, 52\,\text{nm}, 1.001\,\mu\text{m})$.}
    \label{fig:fig3a}
\end{figure}

Now, to determine the solution of the optimization problem among these possible solutions, we calculate for each of them full capacitance gradient profiles at the tip sample distances of the input data, compare them with the input data (\textbf{Fig. 3h}), and calculate the capacitance gradient profile errors for the different distances of the given possible solution (\textbf{Fig. 3i}). From each set of error profiles for a given possible solution, we calculate the cumulative normalized standard deviation, $S$, (see Materials and Methods). $S$ is the parameter that measures how well the calculated capacitance gradient profiles reproduce the corresponding input profiles. For instance, for the profile errors in \textbf{Fig. 3i} corresponding to the possible solution ($\varepsilon_{r,p}$=16.3, $t$=40.6 nm, $l$=991 nm) we obtain $S$=0.05 zF/nm. This procedure is repeated for all the possible solutions identified earlier, ($\varepsilon_{r,p}\left(l,z\right), t\left(l,z\right), l$) in \textbf{Fig. 3g}. The solution to the optimization problem ($\varepsilon_{r,p}^{sol}, t^{sol}, l^{sol}$) corresponds to the possible solution that minimizes the value of $S$. \textbf{Figures 3j, 3k, 3l} show, respectively, plots of $S$ as a function of $\varepsilon_{r,p}$, $t$, and $l$ corresponding to the possible solutions. We observe that, indeed, $S$ presents a minimum, and only one minimum, from which the solution of the inverse problem can be determined. In the present case, we obtain ($\varepsilon_{r,p}^{sol}, t^{sol}, l^{sol}$)=(9.8, 52 nm, 1.001 µm). This solution nicely agrees with the input parameters used to calculate the input EFM images, i.e., ($\varepsilon_{r,p}^{*}, t^{*}, l^{*}$)=(10, 50 nm, 1.000 µm).

Similar conclusions are obtained for the case that the side-length of the buried pillar is smaller than the tip radius. We show it in Fig. 4 where we consider the case of a small buried pillar with ($\varepsilon_{r,p}^{*}, t^{*}, l^{*}$) =(10, 50 nm, 100 nm) (see the input data in this case in \textbf{Figs. 4a, 4b} and \textbf{4c}). By following the same steps as those detailed above (see \textbf{Figs. 4d-4l}), we obtain the solution ($\varepsilon_{r,p}^{sol}, t^{sol}, l^{sol}$)=(9.9, 50.3 nm, 100.3 nm), again very close to the parameters defining the buried structure and used to calculate the input EFM images.

These results demonstrate that the optimization problem posed has a unique solution for ($\varepsilon_{r,p}, t, l$) and that this solution corresponds to the parameters defining the buried structure and used to calculate the input EFM images, ($\varepsilon_{r,p}^{*}, t^{*}, l^{*}$). Therefore, geometric and dielectric information from a buried object can be obtained from EFM images containing sub-surface information, which constitutes the main result of the present work.

\begin{figure}[ht!]
    \centering
    \includegraphics[width=\textwidth]{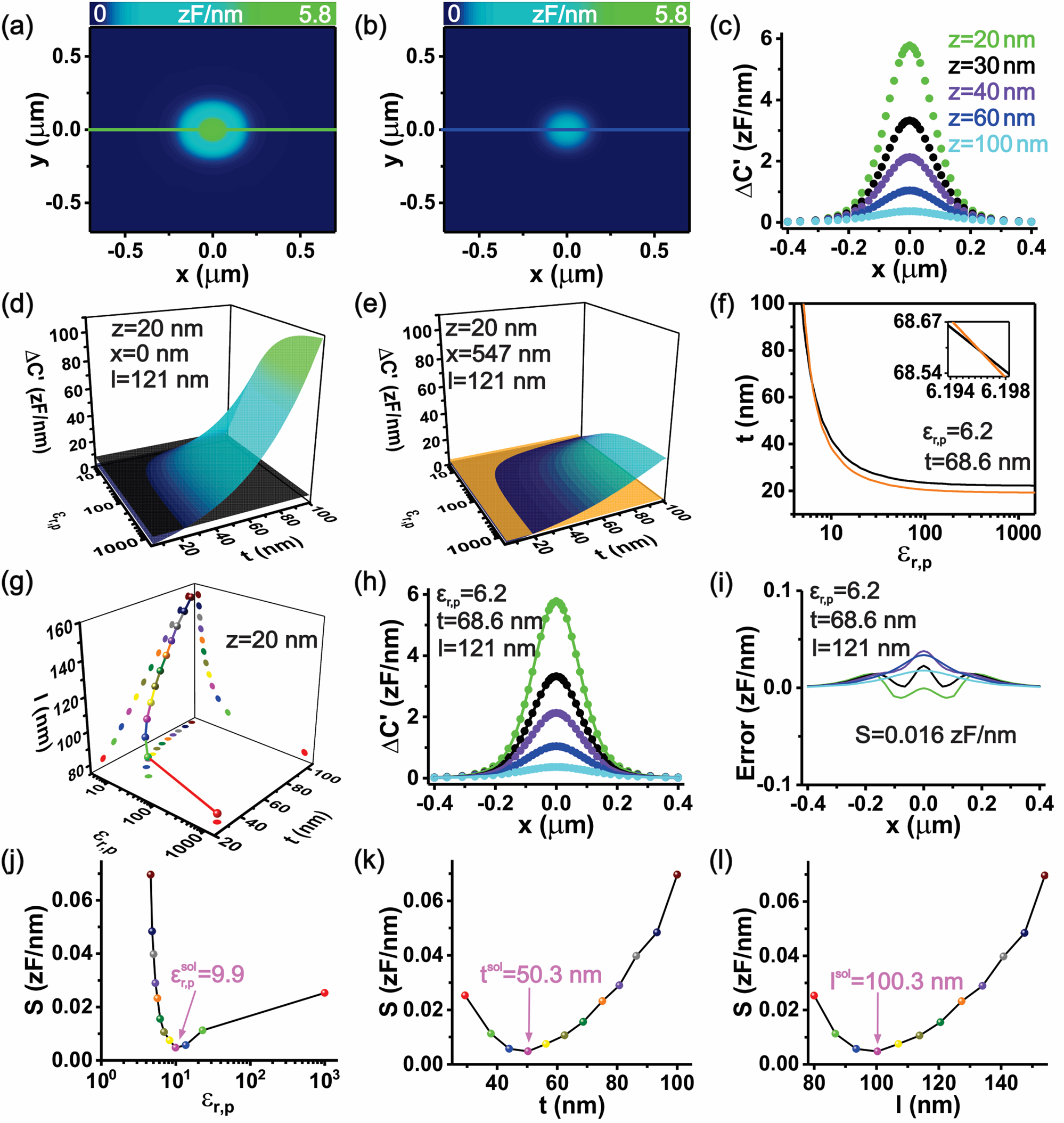}
    \caption{\textbf{(a)} and \textbf{(b)} Examples of input calculated EFM images at tip-sample distances $z = \SI{20}{nm}$ and $z = \SI{60}{nm}$, respectively. Parameters of the buried pillar: thickness, $t^{*} = \SI{50}{nm}$, relative dielectric constant $\varepsilon_{r,p}^{*} = 10$ and side length, $l^{*} = \SI{100}{nm}$. \textbf{(c)} Capacitance gradient cross-section profiles obtained from the input EFM images across the center for $z = \SI{20}{nm}$, \SI{30}{nm}, \SI{40}{nm}, \SI{60}{nm}, \SI{100}{nm}. \textbf{(d)} Calculated possible capacitance gradient contrast values at the center of the buried structure (green surface) as a function of $\varepsilon_{r,p}$ and $t$ for a tip sample distance $z = \SI{20}{nm}$ and a side length of the buried structure $l = \SI{121}{nm}$. The gray plane corresponds to the contrast in the input capacitance gradient profile at the center of the buried structure at this distance. \textbf{(e)} Idem but at the position of the $FW1/3M$. \textbf{(f)} Curves of solutions resulting from the intersection of the surfaces in \textbf{(c)} and \textbf{(d)}, respectively. Inset: zoom in around the position where the curves cross, giving the possible solution value for $l = \SI{121}{nm}$ and $z = \SI{20}{nm}$, $\varepsilon_{r,p} = 6.2$, and $t = \SI{68.6}{nm}$. \textbf{(g)} Plot of the possible solutions for the different set of the parameters that adjust correctly the contrast at the center and at the $FW1/3M$ of the input profile for $z = \SI{20}{nm}$. \textbf{(h)} Comparison of the calculated profiles for one of the possible solutions in \textbf{(f)} (lines) with the input capacitance gradient profiles. \textbf{(i)} Profile capacitance gradient errors resulting from the comparison in (h). The overall error $S$ for this distance and possible solution is $S = \SI{0.016}{zF/nm}$. \textbf{(j)}, \textbf{(k)} and \textbf{(l)} representation of the overall error $S$ as a function of the dielectric constant, thickness and side length of the buried structure, respectively. The minima of the representations correspond to the solution of the problem, which in the present case is ($\varepsilon_{r,p}^{sol}, t^{sol}, l^{sol}$) = (\SI{9.9}{}, \SI{50.3}{nm}, \SI{100.3}{nm}).
}
    \label{fig:fig4}
\end{figure}

\section{Discussion}

We have presented a computational approach to demonstrate the tomographic reconstruction capabilities of electrostatic force microscopy (EFM). To this end we have considered the case of a buried square pillar of thickness $t$, side length $l$, and dielectric constant $\varepsilon_{r,p}$, buried in a dielectric thin film matrix of thickness $h$ and dielectric constant $\varepsilon_{r,m}$. We have shown that, under reasonable assumptions, and by using a specifically posed optimization problem and resolution algorithm, it is possible to obtain from EFM images at different heights quantitative information on three parameters defining the geometry and dielectric constant of the buried pillar ($\varepsilon_{r,p}, t, l$). This result shows that, at least theoretically, a tomographic technique based on EFM can be developed.

On the theoretical side, we note that we have shown the uniqueness of the solution of the optimization problem (Eqs. (1)-(6) in the Materials and Methods) numerically. Demonstrating this result analytically, if possible, lies outside the scope of the present work. At present, we can state that a unique solution for the optimization problem can be found, at least, for a broad range of parameters covering many situations of interest.

On the practical side, it is necessary to evaluate the effect of the experimental noise on the tomographic reconstruction process. The presence of noise in the input data, $\delta C'$, can reduce severely the accuracy with which the different parameters can be extracted, and even can break the possibility to obtain a unique solution. We show it graphically on what follows. The introduction of noise into the input data introduces some uncertainty in the position of the surface planes in \textbf{Figs. 3d} and \textbf{3e}, used to determine the sets of values ($\varepsilon_{r,p}(l, z), t(l, z)$) for given $l$ and $z$ corresponding to imposing the conditions that the input contrast at the center and at the FW1/3M equals one of the possible calculated contrasts (in a sense the noise introduces a "thickness" into the input capacitance gradient planes in \textbf{Figs. 3d} and \textbf{3e} or in \textbf{Figs. 4d} and \textbf{4e}). As a result, the corresponding curves of values (\textbf{Fig. 3f} or \textbf{Fig. 4f}), will display some uncertainty ("thickness"), which will make them to not intersect anymore at a single point but to overlap over a given range of values of the parameters. The range of the overlap dictates the uncertainty with which the possible solutions ($\varepsilon_{r,p}(l, z), t(l, z), l$) can be determined for the given noise. \textbf{Figures 5a} and \textbf{6a} show graphically the overlap ranges between the two curves mentioned, plotted on top of one of the curves, for the case of $z = \SI{20}{nm}$ and $l = \SI{1}{\micro m}$ and $l = \SI{100}{nm}$, respectively, and for different levels of the noise. As it can be seen, the overlap ranges depend strongly on the noise and on the size of the buried object. The overlaps are negligible for noise levels below $\sim \SI{0.1}{zF/nm}$ for $l = \SI{1}{\mu m}$ and below $\sim \SI{0.01}{zF/nm}$ for $l = \SI{100}{nm}$, but they increases quickly as the noise level increases, reaching the whole range of values of the parameters for noise levels of just $\sim \SI{0.5}{zF/nm}$-$\SI{1}{zF/nm}$. The overlap ranges converted into uncertainties in the parameters ($\varepsilon_{r,p}, t$) as a function of the noise level are shown in \textbf{Figs. 5b} and \textbf{5c} for $l = \SI{1}{\micro m}$ and in \textbf{Figs. 6b} and \textbf{6c} for $l = \SI{100}{nm}$. Tolerable uncertainties below a 10\% (gray bands in the figures) can be achieved only for noise levels up to $\sim \SI{0.1}{zF/nm}$ for $l = \SI{1}{\micro m}$ and up to $\sim \SI{0.01}{zF/nm}$ for $l = \SI{100}{nm}$, while for larger noise levels the uncertainty increases beyond these limits quickly.

In the less restrictive case that some additional information on the buried nano-object is known, the sensitivity requirements are less demanding. For instance, if one considers that the dielectric constant of the buried material is known, information on the geometric dimensions of the buried pillar (thickness and side length) can be obtained with an error below a 10\% for noise levels ranging from \(\sim\)1 zF/nm for \(l = 1~\mu\)m to \(\sim\)0.1 zF/nm for \(l = 100\) nm (see \textbf{Figs.~5d--5f} and \textbf{Figs.~6d--6f}, respectively). A similar, but slightly worse, situation is found if the additional known parameter corresponds to the thickness, \(t\), (see \textbf{Figs.~5g--5i} and \textbf{Figs.~6g--6i}) or to the lateral side, \(l\) (see Figs.~5a--5c and Figs.~6a--6c), especially for small-scale pillars, where noise levels below \(\sim\)0.1 zF/nm can still be necessary.

Finally, in the case that only one parameter of the buried structure remains unknown, accurate values for this parameter can be obtained with an error below 10\% for noise levels up to \(\sim\)1 zF/nm, even for the smallest structure (see \textbf{Figs.~5j} and \textbf{6j} for the case where the dielectric constant is the single unknown parameter, \textbf{Figs.~5k} and \textbf{6k} when it is the thickness, and \textbf{Figs.~5l} and \textbf{6l} when it is the side length). An experimental example of the latter case has been shown recently for water confined in nanochannels~\cite{fumagalli2018}, where the dielectric constant of the confined water was accurately determined for a known geometry of the buried nanochannel.

The detection noise of EFM set ups covers a broad range of values between \(\sim\)0.01--0.1 zF/nm\cite{fumagalli2012} up to \(\sim\)1--5 zF/nm, depending on the characteristics of the probe (e.g.\ spring constant) and of the detection system (e.g.\ photodiode, control and acquisition electronics, etc.). Therefore, with the use of fine tuned EFM systems, tomographic EFM reconstruction should be possible in practice, although challenging, specially for small scale objects or objects buried deep into the surrounding matrix (beyond 100 nm typically). To this respect, it should be noted that additional sources of uncertainty could influence negatively the implementation of this SPM tomographic technique, such as the accuracy of the tip geometry calibration or the determination of the tip-sample distance, but it is expected that they would not change significantly the conclusion obtained above. Finally, in the case of non-planar samples further complexities are expected, whose discussion lies outside the scope of the present work (see for instance~\cite{van2016,van2016a}).

\begin{figure}[htbp]
    \centering
    \includegraphics[width=\textwidth]{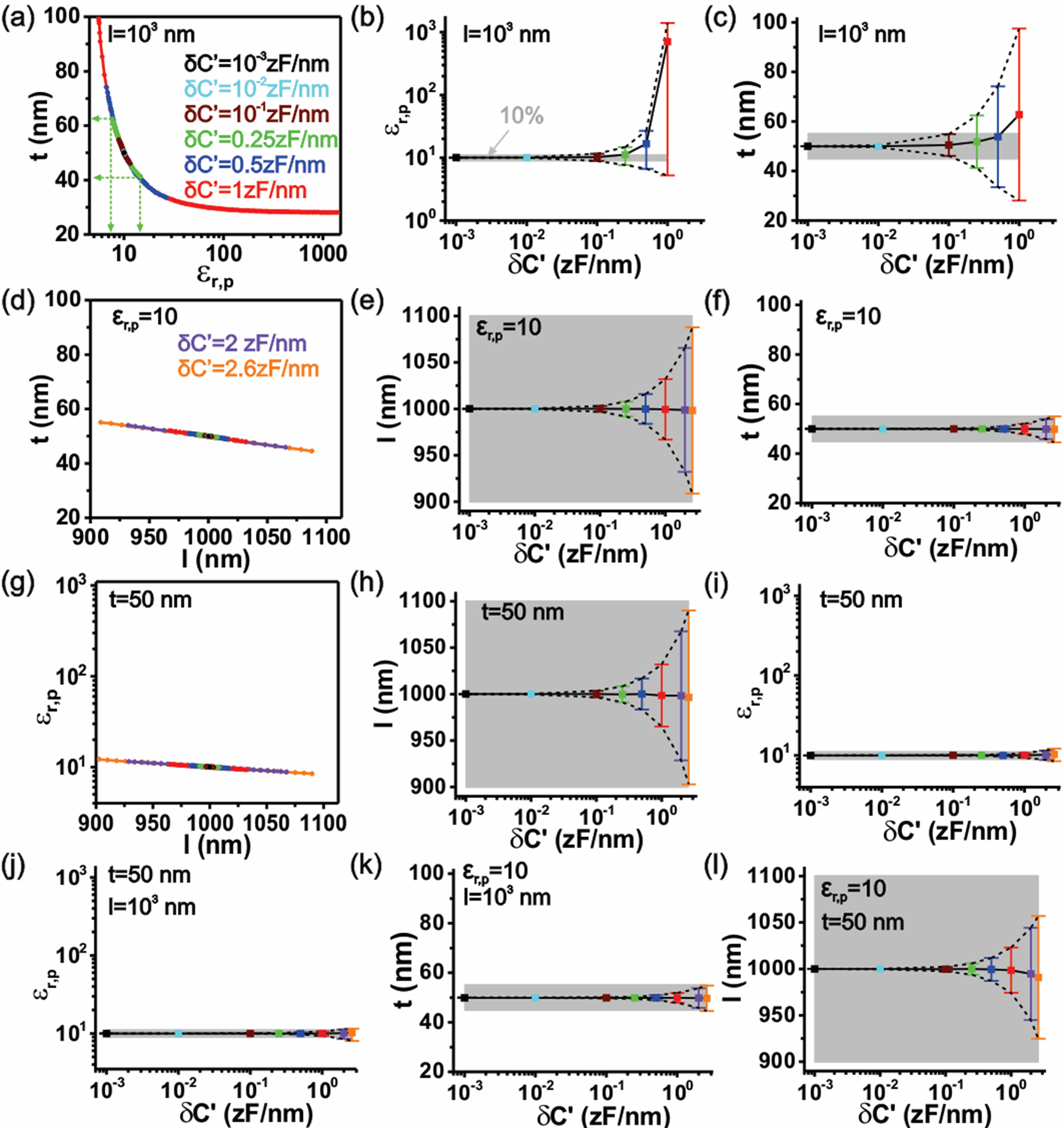}
    \caption{\textbf{(a)} Graphical representation of the overlap ranges between curves of the type shown in \textbf{Fig.~3c}, calculated from the contrast at $x=0$ and $x=FW1/3M$ as a function of the noise in the input EFM data, for $l = \SI{1}{\micro m}$ and $z = \SI{20}{nm}$. \textbf{(b)} and \textbf{(c)} Uncertainty in the dielectric constant and the pillar thickness, respectively, as a function of the input noise. The gray band indicates the 10\% relative error limit. \textbf{(d)} Same as (a) but for the thickness and the side length of the buried pillar for a given dielectric constant, $\varepsilon_{r,p} = 10$. \textbf{(e)} and \textbf{(f)} Same as (b) and (c) but for the side length and thickness, respectively. \textbf{(g)} Same as (a) but for the dielectric constant and the side length for a given thickness $t = \SI{50}{nm}$. \textbf{(h)} and \textbf{(i)} Same as (b) and (c) but for the side length and the dielectric constant. \textbf{(j)} Same as (b) but for a given thickness $t = \SI{50}{nm}$ and side length $l = \SI{1}{\micro m}$. \textbf{(k)} Same as (j) but for the pillar thickness for $\varepsilon_{r,p} = 10$ and $l = \SI{1}{\micro m}$. \textbf{(l)} Same as (j) but for the pillar side length with $\varepsilon_{r,p} = 10$ and $t = \SI{50}{nm}$. Parameters: same as in Figure \ref{fig:fig3a}, unless otherwise indicated.}
    \label{fig:fig5a}
\end{figure}

\begin{figure}[H]
    \centering
    \includegraphics[width=\textwidth]{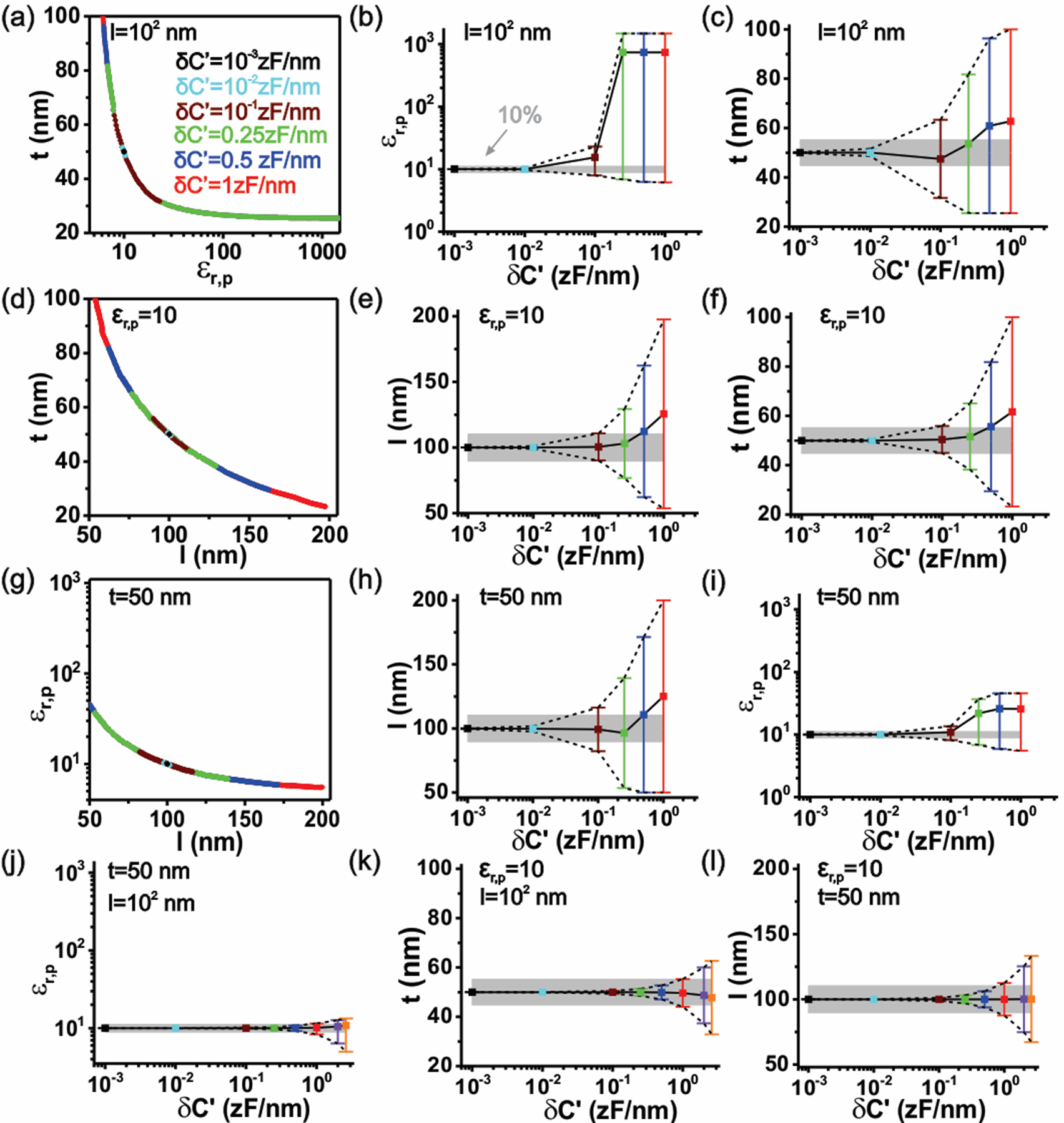}
   \caption{\textbf{(a)} Graphical representation of the overlap ranges between curves of the type shown in \textbf{Fig.~3f}, calculated from the contrast at $x=0$ and $x=FW1/3M$ as a function of the noise in the input EFM data, for $l = \SI{100}{nm}$ and $z = \SI{20}{nm}$. \textbf{(b)} and \textbf{(c)} Uncertainty in the dielectric constant and the pillar thickness, respectively, as a function of the input noise. The gray band indicates the 10\% relative error limit. \textbf{(d)} Same as (a) but for the thickness and the side length of the buried pillar for a given dielectric constant, $\varepsilon_{r,p} = 10$. \textbf{(e)} and \textbf{(f)} Same as (b) and (c) but for the side length and thickness, respectively. \textbf{(g)} Same as (a) but for the dielectric constant and the side length for a given thickness $t = \SI{50}{nm}$. \textbf{(h)} and \textbf{(i)} Same as (b) and (c) but for the side length and the dielectric constant. \textbf{(j)} Same as (b) but for a given thickness $t = \SI{50}{nm}$ and side length $l = \SI{100}{nm}$. \textbf{(k)} Same as (j) but for the pillar thickness for $\varepsilon_{r,p} = 10$ and $l = \SI{100}{nm}$. \textbf{(l)} Same as (j) but for the pillar side length with $\varepsilon_{r,p} = 10$ and $t = \SI{50}{nm}$. Parameters: same as in Figure \ref{fig:fig4}, unless otherwise indicated.}
    \label{fig:fig6a}
    \vskip-.25cm
\end{figure}

\section{Materials and methods.}

\subsection{Finite Element numerical calculations.}

We used 3D finite element numerical calculations to compute Electrostatic Force Microscopy images following earlier 3D works~\citep{fumagalli2018,van2016,van2016a,lozano2018}. A schematic representation of the system simulated is shown in Fig.~1a. The tip was represented like in previous works~\cite{fumagalli2018a} by a metallic truncated cone of height \(H\), and half-cone angle \(\theta\), ended with a tangent sphere of radius \(R\), and covered by a disc of thickness \(w\), and radius \(R \cdot \tan(\theta) + L\), representing local cantilever effects. Nonlocal cantilever contributions, if necessary, can be considered through a phenomenological capacitance gradient offset contribution term~\cite{gramse2012}. 

The simulation domain was cylindrical with radius \SIrange{16}{25}{\mu m} and a height \SIrange{30}{50}{\mu m}, depending on the tip position with respect to the sample. The mesh was set to a minimum of \(6 \times 10^{5}\) elements (see further details in~\cite{lozano2018}) in order to provide the convergence of the results. An accurate process of optimization, validation, mesh convergence and numerical noise reduction of the 3D simulations has been undertaken, to meet sub-\SI{0.1}{zF/nm} numerical noise accuracy (\(<\!10^{-3}\)~\si{zF/nm}, see further details in Refs.~\cite{lozano2018} and~\cite{wu2017}). 

The electric potential distribution between the tip and the sample was calculated by solving Poisson's equation with the AC/DC electrostatic module of Comsol Multiphysics 5.3a linked to MATLAB. We set the surface of the tip to ``terminal'', the bottom boundary of the simulation domain to ``ground'', and the top and side boundaries to ``zero charge''. The infinite element function was used on the top and side boundaries to get rid of finite size effects of the simulation domain. We calculated the electric force on the tip by integration on the tip surface of the built-in Maxwell stress tensor (see further details in Ref.~\cite{fumagalli2018a}).

\subsection{Standard form of the optimization problem.}

The standard form of the optimization problem solved to determine the unknown parameters describing the buried pillar \(\left(\varepsilon_{r,p}, t, l\right)\) can be formulated as follows:

\begin{equation}
    \min S\left(\varepsilon_{r,p}, t, l\right)
    \label{eq:main_opt}
\end{equation}

subject to

\begin{align}
    &\left| \Delta C^{\prime}(\hat{x}_{0}; \varepsilon_{r,p}, t, l) - \Delta C^{\prime}_{in}(\hat{x}_{0}) \right| \leq \delta C^{\prime}
    \label{eq:constraint1} \\
    &\left| \Delta C^{\prime}(\hat{x}_{1}; \varepsilon_{r,p}, t, l) - \Delta C^{\prime}_{in}(\hat{x}_{1}^{\flat}) \right| \leq \delta C^{\prime}
    \label{eq:constraint2} \\
    &1 \leq \varepsilon_{r,p} < +\infty
    \label{eq:constraint3} \\
    &0 < t \leq h
    \label{eq:constraint4} \\
    &0 < l < +\infty
    \label{eq:constraint5}
\end{align}

where we denote the tip position by \(\hat{x} = (x, y, z)\), the experimental noise by \(\delta C^{\prime}\), the capacitance gradient contrast by \(\Delta C^{\prime}\), and the cumulative standard deviation along a profile \(S\) by:

\begin{equation}
    S = \left( \frac{1}{X_{T}Z_{T}} \int_{-Z_{T}/2}^{Z_{T}/2} \int_{-X_{T}/2}^{X_{T}/2} E_{r}^{2}\,(\hat{x}) \mathrm{d}x \, \mathrm{d}z \right)^{1/2}.
    \label{eq:cumulative_std}
\end{equation}

Here \(E_{r}(\hat{x}) = \Delta C^{\prime}(\hat{x}; \varepsilon_{r,p}, t, l) - \Delta C^{\prime}_{in}(\hat{x})\) denotes the profile error, \(X_{T}\) the length of the input capacitance gradient profile, and \(Z_{T}\) the tip-sample distance range covered. Moreover, \(\Delta C^{\prime}_{in}(\hat{x})\) is the input capacitance gradient at the tip position \(\hat{x} = (x, y, z)\), and \(\Delta C^{\prime}(\hat{x}; \varepsilon_{r,p}, t, l)\) the computed capacitance gradient image for given values of the unknown parameters \((\varepsilon_{r,p}, t, l)\). The positions \(\hat{x}_{0}\) and \(\hat{x}_{1}\) correspond to the center and FW1/3M of the EFM image. The defined optimization problem should be understood as a numerical minimization problem with an accuracy within the numerical error of the numerical calculations (in the present case \(<\!10^{-3}~\si{zF/nm}\)).

\section{Conclusions.}

We have shown theoretically that the sub-surface imaging capability of Electrostatic Force Microscopy can be used to extract tomographic information (physical dimensions and dielectric constant) of buried nanostructures in thin dielectric films. To show it we have presented an optimization problem, whose numerical resolution by means of a developed algorithm, has given a unique solution, which corresponds to the parameters of the buried structure. The sensitivity of the tomographic reconstruction algorithm to the presence of noise in the input EFM data has been evaluated. It has been concluded that high sensitivity EFM instruments (noise levels below $\sim \SI{0.1}{zF/nm}$, especially for small scale objects below $\sim \SI{100}{nm}$ size) might be required for a practical and optimal implementation of the technique. The results of the present work have been developed for the case of a square parallelepiped nanostructure buried within a thin dielectric matrix, but it can be easily extended to other geometries of the buried objects. Present results constitute a first step towards implementing a non-destructive dielectric nanotomography technique based on EFM with applications in Materials and Life Sciences.

\section{Acknowledgments.}

This work has been partially supported by the Spanish Agencia Estatal de Investigación and EU FEDER through grants TEC2013-48344-C2-1-P and TEC2016-79156-P. We also acknowledge support from Generalitat de Catalunya through 2017-SGR1079 grant, CERCA Program and ICREA Academia Award (G.G.). This work has received funding from the European Union's Horizon 2020 research and innovation program under the Marie Skłodowska-Curie grant agreement No. 721874 (SPM2.0). We acknowledge Dr. Laura Fumagalli for fruitful discussions on the topic.

\newpage
\bibliographystyle{unsrtnat} 
\bibliography{references}

\end{document}